\documentclass{icrc2009}

\usepackage[footnotesize]{caption}
\usepackage{graphicx}   
\usepackage[font=footnotesize]{subfig} 
\usepackage{fixltx2e}
\usepackage{url}

\newcommand{\shorttitle}[1]%
{\markboth{Proceedings of the 31\MakeLowercase{$^{st}$} ICRC, {\L}\'{o}d\'{z} 2009}{#1} }
\newcommand{\etal}{\MakeLowercase{\textit{et al. }}} 

\newcommand{\un}[1]{\,\mathrm{#1}}

\begin{document}

\baselineskip 11.8pt

\title{Supernova Search with the AMANDA / IceCube Detectors}

\author{
\IEEEauthorblockN{
Thomas Kowarik\IEEEauthorrefmark{1},
Timo Griesel\IEEEauthorrefmark{1},
Alexander Pi\'egsa\IEEEauthorrefmark{1}
for the IceCube Collaboration\IEEEauthorrefmark{2}
}\\
\IEEEauthorblockA{\IEEEauthorrefmark{1} Institute of Physics, University of Mainz, Staudinger Weg 7, D-55099 Mainz, Germany}
\IEEEauthorblockA{\IEEEauthorrefmark{2} http://www.icecube.wisc.edu/collaboration/authorlists/2009/4.html} 
}

\shorttitle{T.Kowarik \etal Supernova Search with IceCube}
\maketitle

\begin{abstract}
Since 1997 the neutrino telescope AMANDA at the geographic South Pole has been monitoring our Galaxy for 
neutrino bursts from supernovae. Triggers were introduced in 2004 to submit burst candidates to the
Supernova Early Warning System SNEWS. From 2007 the burst search was extended to the much larger IceCube
telescope, which now supersedes AMANDA. By exploiting the low photomultiplier noise in the antarctic ice
(on average 280\,Hz for IceCube), neutrino bursts from nearby supernovae can be identified by the induced 
collective rise in the pulse rates. Although only a counting experiment, IceCube will provide the world's 
most precise measurement of the time profile of a neutrino burst near the galactic center. The sensitivity 
to neutrino properties such as the $\theta_{13}$ mixing angle and the neutrino hierarchy are discussed as 
well as the possibility to detect the deleptonization burst.
\end{abstract}

\begin{IEEEkeywords}
	supernova neutrino IceCube 
\end{IEEEkeywords}

\section{Introduction}

Up to now, the only detected extra-terrestrial sources of neutrinos are the sun and supernova SN1987A. To
extend the search to TeV energies and above, neutrino telescopes such as AMANDA and IceCube 
\cite{i3description} have been built. It turns out that the noise rates of the light sensors (OMs) in the
antarctic ice are very low ($\sim 700\un{Hz}$ for AMANDA, $\sim280\un{Hz}$ for IceCube) opening up the
possibility to detect  MeV electron anti-neutrinos from close supernovae by an increase in the collective
rate of all light sensors. The possibility to monitor the galaxy for supernova with neutrino telescopes
such as AMANDA has first been proposed in~\cite{Halzen1994} and a first search has been performed using data
from the years 1997 and 1998 \cite{firstAMANDAsnsearch}.

The version of the the supernova data acquisition (SNDAq) covered in this paper has been introduced for AMANDA
in the beginning of the year 2000 and was extended to IceCube in 2007. The AMANDA SNDAq has been switched off
in February 2009. We will investigate data recorded by both telescopes concentrating on the 9 years of AMANDA
measurements and make predictions for the expected sensitivity of IceCube.

\section{Detectors and Data Acquisition}

In AMANDA, the pulses of the 677 OMs are collected in a VME/Linux based data acquisition system which
operates independently of the main data  acquisition aimed at high energy neutrinos. It counts pulses
from every connected optical module in a 20 bit counter in fixed $10\un{ms}$ time intervals that are 
synchronized by a GPS-clock.

In IceCube, PMT rates are recorded in a $1.6384 \un{ms}$ binning by scalers on each optical module. The
information is locally buffered and read out by the IceCube data acquisition system. It then transfers
this data to the SNDAq, which synchronizes and regroups the information in 
$2 \un{ms}$ bins. 

The software used for data acquisition and analysis is essentially the same for AMANDA and IceCube.
The data is rebinned in $500 \un{ms}$ intervals and subjected to an online analysis described later.
In case of a significant rate increase (``supernova trigger''), an alarm is sent to the Supernova Early
Warning System (SNEWS, \cite{snews}) via the Iridium satellite network and the data is saved in a fine
time binning ($10\un{ms}$ for AMANDA and $2\un{ms}$ for IceCube).

\section{Sensor Rates}

\begin{figure}
	\begin{center}
		\includegraphics [width=0.47\textwidth]{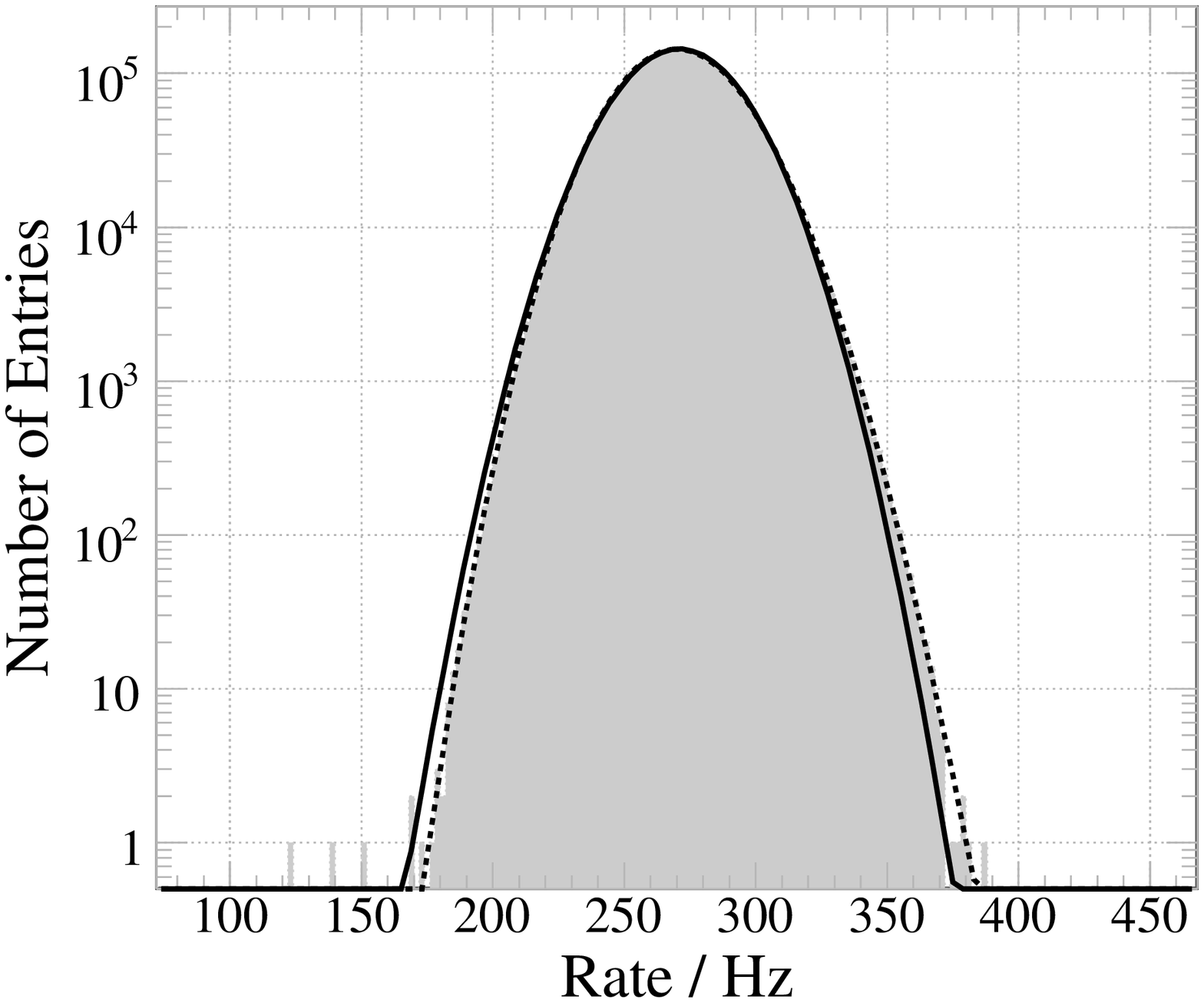}
	\end{center}
	\vspace{-0.5cm}
	\caption{Rate distribution of a typical IceCube module}
	\footnotesize{The recorded rate distribution (filled area) encompasses about 44 days and has
	been fitted with a Gaussian (solid line, $\chi^2 / N_\mathrm{dof}=56.5$) defined by $\mu=271\un{Hz}$
	and $\sigma=21\un{Hz}$. However, a lognormal function (dotted line, $\chi^2 / N_\mathrm{dof}=0.66$) 
	with geometric mean \mbox{$\mu_\mathrm{geo}=6.4\un{\ln(Hz)}$}, geometric standard deviation 
	$\sigma_\mathrm{geo}=0.03\un{\ln(Hz)}$ 	and a shift of $x_0=377\un{Hz}$ is much better at describing 
	the data.}	
	\label{fig::dom_lognormal_fit}
\end{figure}

In $500 \un{ms}$ time binning, the pulse distribution of the average AMANDA or IceCube OM conforms only 
approximately to a Gaussian. It can more accurately be described by a lognormal distribution (see
figure~\ref{fig::dom_lognormal_fit}).

The pulse distributions exhibit Poissonian and correlated afterpulse components contributing with similar
strengths. The correlated component is anticorrelated with temperature and arises form Cherenkov light
caused by $^{40}\mathrm{K}$ decays and glass luminescence from radioactive decay chains. A cut at
$250 \un{\mu s}$ on the time difference of consecutive pulses effectively suppresses afterpulse trains,
improves the significance of a simulated supernova at $7.5\un{kpc}$ by approximately $20\%$ and
makes the pulse distribution more Poissonian in nature. 

\section{Effective Volumes}

Supernovae radiate all neutrino flavors, but due to the relatively large inverse beta decay cross section,
the main signal in IceCube is induced by electron anti-neutrinos (see~\cite{Haxton1987}). The rate $R$
per OM can be approximated weighing the energy dependent anti-electron neutrino flux at the earth 
$\Phi(E_{\bar\nu_e})$  (derived from the neutrino luminosity and spectra found in~\cite{Keil2003}) with the 
effective area for anti-electron neutrino detection $A_\mathrm{eff}(E_{\bar\nu_e})$ and integrating over 
the whole energy range:

\begin{eqnarray*}
&\hspace{-0.3cm} R = \int^{\infty}_0 \! dE_{\bar\nu_e} \, \Phi(E_{\bar\nu_e}) A_\mathrm{eff}(E_{\bar\nu_e})
\;,\; \mathrm{with}\\
&\hspace{-0.3cm} A_\mathrm{eff}(E_{\bar\nu_e}) = n \int^{\infty}_{0} \! dE_{e^+} \; 
	\frac{d\sigma}{dE_{e^+}}(E_{\bar\nu_e}, E_{e^+}) \, V_{\mathrm{eff},e^+}(E_{e^+})\,.
\end{eqnarray*}

\noindent $\frac{d\sigma}{dE_{e^+}}(E_{\bar\nu_e}, E_{e^+})$ is the inverse beta decay cross section, $n$
the density of protons in the ice and $V_{\mathrm{eff}, e^+}(E_{e^+})$ the effective volume for positron 
detection of a single OM.

$V_{\mathrm{eff}, e^+}(E_{e^+})$ can be calculated by multiplying the number of Cherenkov photons produced 
with the effective volume for photon detection $V_\mathrm{eff,\gamma_\mathrm{ch}}$. 

By tracking Cherenkov photons in the antarctic ice around the IceCube light sensors~\cite{Lundberg2007}
and simulating the module response one obtains $V_\mathrm{eff,\gamma_\mathrm{ch}}= 0.104\un{m^3}$ 
for the most common AMANDA sensors and $V_\mathrm{eff,\gamma}= 0.182\un{m^3}$ for the IceCube sensors. 
With a \mbox{\emph{GEANT-4}} simulation, the amount of photons produced by a positron of an energy $E_{e^+}$ 
can is estimated to be $N^{\gamma_\mathrm{ch}} = 270 \, E_{e^+}/\un{MeV}$. Consequently, the 
effective volumes for positrons as a function of their energies are 
$V^\mathrm{AMANDA}_{\mathrm{eff}, e^+}= 19.5 \, E_{e^+}\un{\frac{m^3}{MeV}}$ and
$V^\mathrm{IceCube}_{\mathrm{eff}, e^+}= 34.2 \, E_{e^+}\un{\frac{m^3}{MeV}}$. Uncertainties in the effective
volumes derive directly from the uncertainties of the ice models ($\sim 5\%$) and in the OM sensitivities
($\sim 10\%$).

In this paper we assume a supernova neutrino production according to the \emph{Lawrence-Livermore}
model~\cite{Totani1998} as it is the only one that provides spectra for up to $15\un{s}$. It gives a mean
electron anti-neutrino energy of $15\un{MeV}$ corresponding to an average positron energy of 
$( 13.4 \pm 0.5 ) \un{MeV}$.

\section{Analysis Procedure}

A simple investigation of the rate sums would be very susceptible to fluctuations due to variations in
the detector response or external influences such as the seasonal variation of muon rates. Medium and
long term fluctuations are tracked by estimating the average count rate by a sliding time window. The
rate deviation for a collective homogeneous neutrino induced ice illumination is calculated by a 
likelihood technique. In time bins of $0.5\un{s}$ and longer, the pulse distributions can be approximated
by Gaussian distributions. For a collective rate increase $\Delta\mu$, the expectation value of the
average mean rate $\mu_i$ of a light sensor $i$ with relative sensitivity $\epsilon_i$ increases to 
$\mu_i + \epsilon_i \Delta\mu$. The mean value $\mu_i$ and its standard deviation $\sigma_i$ are averaged
over a sliding window of $10\un{min}$, excluding $15\un{s}$ before and after the $0.5 \un{s}$ time frame
$r_i$ studied. By taking the product of the corresponding Gaussian distributions the following likelihood 
for a rate deviation $\Delta\mu$ is obtained:

\begin{displaymath}
	{\cal L} = \prod_{i=1}^{N_\mathrm{OM}}\, \frac{1}{{\sqrt{2\pi}\,\sigma_i}} 
		\exp{-\frac{(r_i-(\mu_i+\epsilon_i\,\Delta\mu))^2}{2\sigma_i^2}}
\quad .
\end{displaymath}

\noindent Minimization of $-\ln{\cal L}$ leads to:
\begin{displaymath}
	\Delta\mu = \underbrace{\left(\sum_{i=1}^{N_\mathrm{OM}} \, 
		\frac{\epsilon_i^2}{\sigma_i^2}\right)^{-1}}_{=\;\sigma^2_{\Delta\mu}} 
		\sum_{i=1}^{N_\mathrm{OM}} \, \frac{\epsilon_i\,(r_i - \mu_i)}{\sigma_i^2}
\quad .
\end{displaymath}

\noindent The data is analyzed in the three time binnings $0.5\un{s}$, $4\un{s}$ and $10\un{s}$ for the following
reasons: First, the finest time binning accessible to the online analysis is $0.5\un{s}$. Second, as argued 
in~\cite{Beacom1999}, the neutrino measurements of SN1987A are roughly compatible with an exponential decay of 
$\tau=3\un{s}$. The optimal time frame for the detection of a signal with such a signature is $\approx 3.8\un{s}$. 
Last, $10\un{s}$ are the approximate time frame where most of the neutrinos from SN1987A fell.

To ensure data quality, the optical modules are subjected to careful quality checks and cleaning. Those
modules with rates outside of a predefined range, a high dispersion w.r.t. the Poissonian expectation or
a large skewness are disqualified in real time. 

Since SNEWS requests one alarm per 10 days, the supernova trigger is set to $6.3\,\sigma$.

To ensure that the observed rate deviation is homogeneous and isotropic, the following $\chi^2$
discriminant is examined:
\begin{displaymath}
	\chi^2(\Delta\mu) = \sum_{i=1}^{N_\mathrm{OM}} 
		\left(\frac{r_i-(\mu_i+\epsilon_i\,\Delta\mu)}{\sigma_i}\right)^2
	\quad .
\end{displaymath}

\noindent We demand the data to conform to a $\chi^2$-confidence level of $99.9\un{\%}$. However, it was 
found that the $\chi^2$ cannot clearly distinguish between isotropic rate changes and fluctuations of a
significant number of OMs: If e.g. $20\%$ of the OMs record rate increases of about  $1\,\sigma$
($\approx 20 \un{pulses/0.5\un{s}}$), the significance for isotropic illumination can rise above $6\,\sigma$
without being rejected by the $\chi^2$ condition. Still, no method was found that performed better than
$\chi^2$.

\section{External Perturbations}

\begin{figure}
	\begin{center}
		\includegraphics [width=0.47\textwidth]{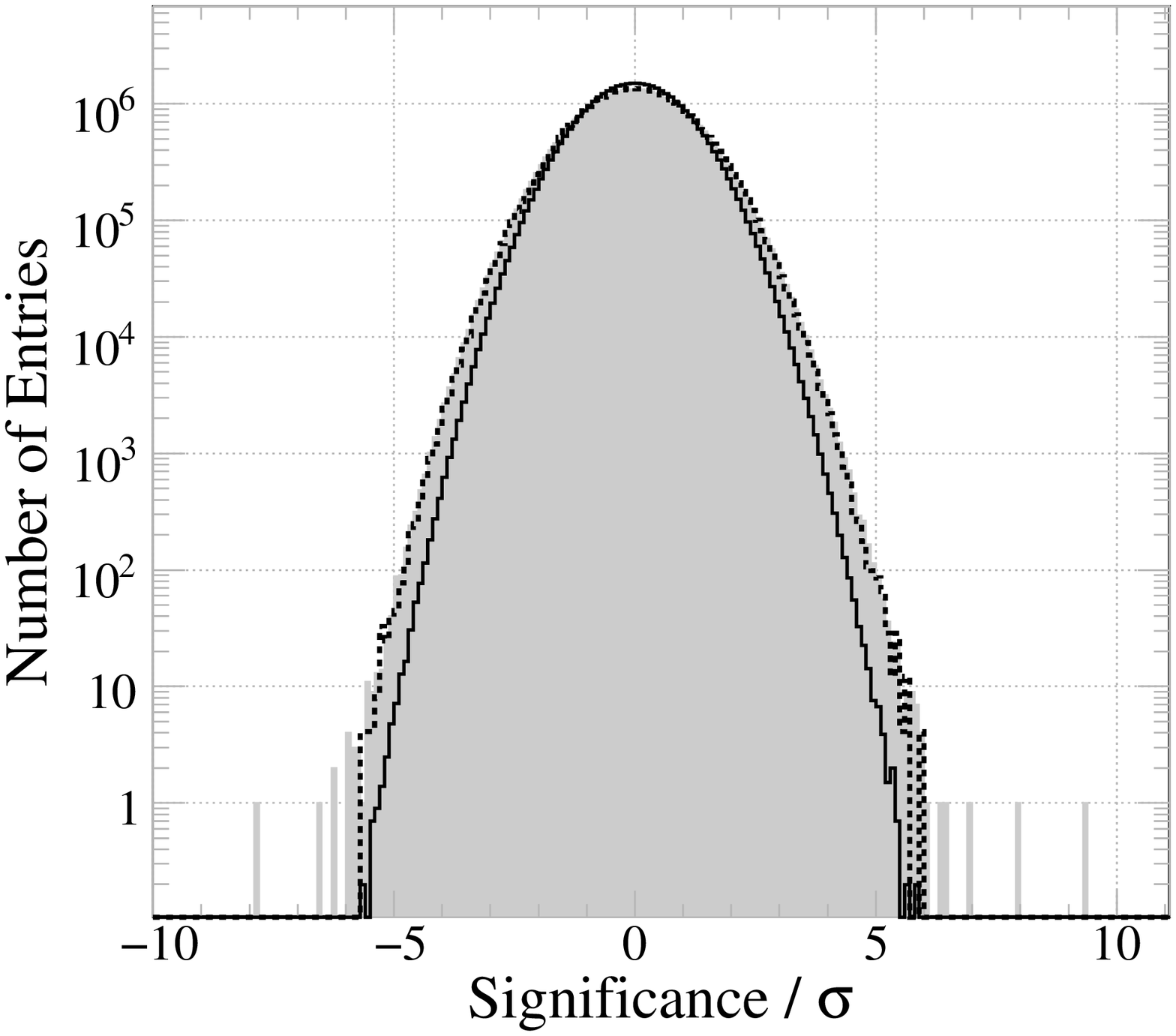}
	\end{center}
	\vspace{-0.3cm}
	\caption{Significances of AMANDA in $0.5\un{s}$ time frames}
	\footnotesize{The filled area shows the significance of the data taken in 2002 with a spread
	of $\sigma=1.13$. The solid line denotes the initial background simulation ($\sigma \approx 1$)
	and the dotted line is a toy monte carlo taking into account the fluctuations due to muon rates
	($\sigma \approx 1.11$).}
	\label{fig::sign_distri}
\end{figure}

Figure~\ref{fig::sign_distri} shows the distribution of the significance
$\xi = \frac{\Delta\mu}{\sigma_{\Delta\mu}}$ for AMANDA. From the central limit theorem, one would expect
a Gaussian distribution with a width of $\sigma = 1$. This expectation is supported by a background
simulation using lognormal representations of individual light sensor pulse distributions. However, one
finds that the observable is spread wider than expected and exhibits a minor shoulder at high
significances.  

As this observable is central to the identification of supernovae, its detailed understanding is imperative
and a close investigation of these effects is necessary.

Faults in the software were ruled out by simulating data at the most basic level, before entering the
SNDAq. Hardware faults are improbable, as the  broadening of the rates is seen both for AMANDA
($\sigma=1.13$) and IceCube ($\sigma=1.27$) and the detectors use different and independent power and
readout electronics.

One can then check for external sources of rate changes as tracked by magnetometers, rio- and photometers
as well as seismometers at Pole. All have been synchronized with the rate measurements of the OMs. Only
magnetic field variations show a slight, albeit insignificant, influence on the rate deviation of 
$-4\cdot 10^{-5}\un{\frac{Hz}{nT}}$ for AMANDA. Due to a $\mu$ metal wire mesh shielding, the influence
on IceCube sensors is smaller by a factor $\sim 30$. 

It turns out that the main reason for the broadening are fluctuations of the atmospheric muon rates. 
During $0.5\un{s}$, AMANDA detects between $\approx 3.0\cdot10^3$ (June) and $\approx 3.4\cdot10^3$ 
(December) sensor hits due to muons. Adding hits from atmospheric muons broadens and distorts the 
distribution derived from noise rates. A simulation taking into account the muon triggered PMT hits 
increases the width of the significance distribution to $\approx 1.11$ (see figure~\ref{fig::sign_distri}).
Although the investigations are still ongoing for IceCube, we expect the larger $\sigma$ can be ascribed 
to its lower noise rate leading to a higher resolution and thereby a stronger sensitivity to perturbations.

A Fourier transformation was performed on the data stream. No evidence for periodically recurring events
was found, but occasional correlations between subsequent $0.5 \un{s}$ time frames both in the summed noise 
rate and the rate deviation could be identified. We determined the mean significance before and after a
rate increase exceeding a predefined level. One observes a symmetric correlation in the data mainly between
$\pm 10\un{s}$. The origin of the effect, which is present both in AMANDA and IceCube data, is under 
investigation.

\section{Expected Signal and Visibility Range}

A preliminary analysis of AMANDA data from 2000 to 2003 yielded a detection range of $R_{4\un{s}}
= 14.5\un{kpc}$ at the optimal binning of $4\un{s}$ at an efficiency of $90\%$, encompassing $81\%$ of
the stars of our Galaxy. As signals with exponentially decreasing luminosity (at $\tau=3\un{s}$) have 
been used as underlying models, the two other binnings were less efficient with $R_{0.5\un{s}} =
10\un{kpc}$ ($56\%$) and $R_{10\un{s}} = 13\un{kpc}$ ($75\%$), respectively. 

The rates of the IceCube sensors are stable and uniform across the detector. Scaling the observed rates
to the full 80 string detector, a summed rate  of $\langle R_\mathrm{IceCube} \rangle = (1.3\cdot10^6 
\pm 1.8\cdot10^2) \un{Hz}$ is expected.  While a single DOM would only see an average rate increase of
$13\un{Hz}$ or $0.65\,\sigma$, the signal in the whole IceCube detector would be $6.1\cdot10^4\un{Hz}$
or $34\,\sigma$ for a supernova at $7.5 \un{kpc}$ distance. Using this simple counting method, IceCube 
would see a supernova in the Magellanic Cloud with $5\,\sigma$ significance.

\begin{figure}[b]
	\begin{center}
		\includegraphics [width=0.44\textwidth]{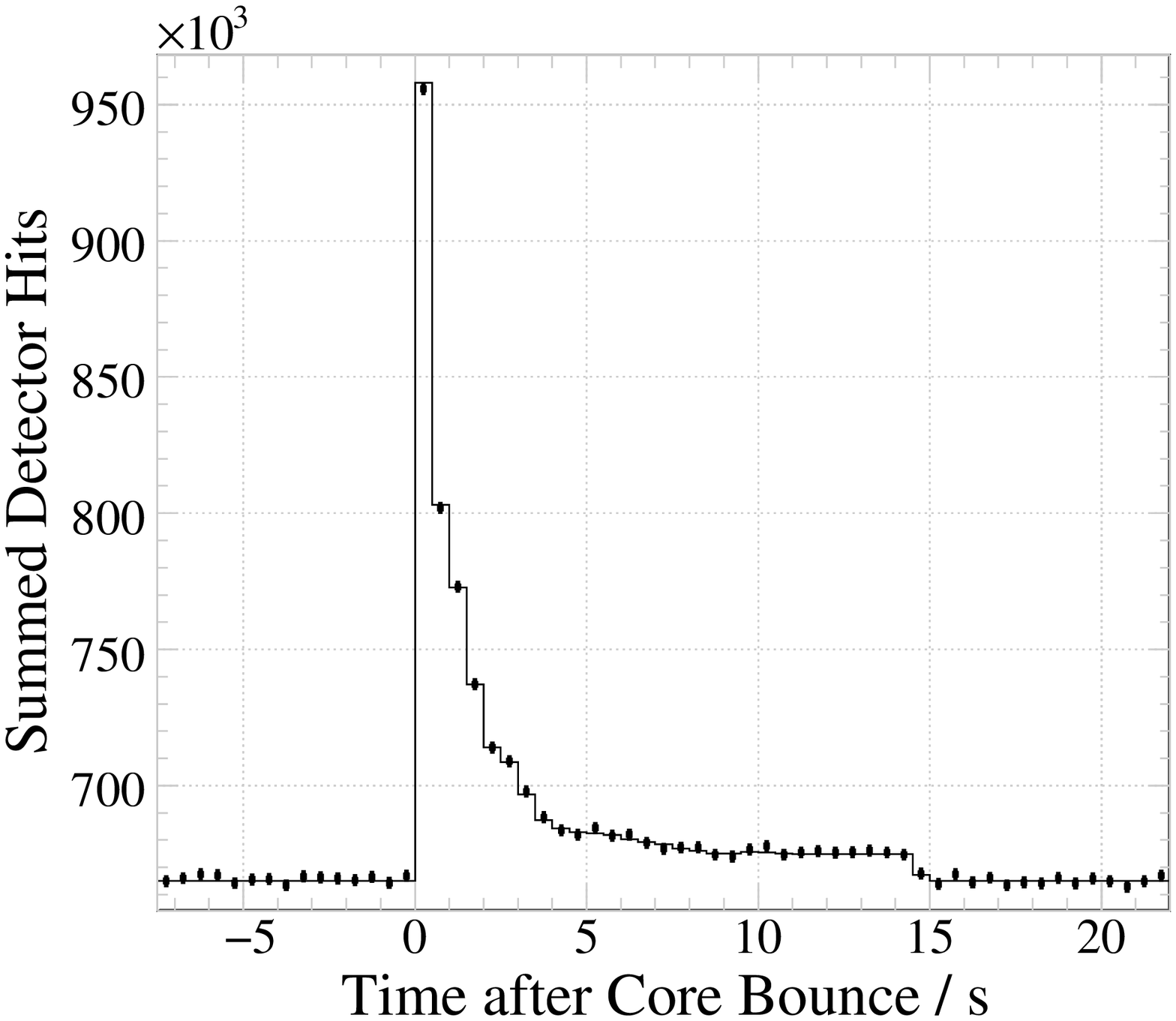}
	\end{center}
	\vspace{-0.3cm}
	\caption{Expected signal of a supernova at $7.5\un{kpc}$ in IceCube.}
	\footnotesize{The solid line denotes the signal expectation and the bars a randomized simulation.}
	\label{fig::ll_sn_7.5kpc}
\end{figure}

\begin{figure}
	\begin{center}
		\includegraphics [width=0.45\textwidth]{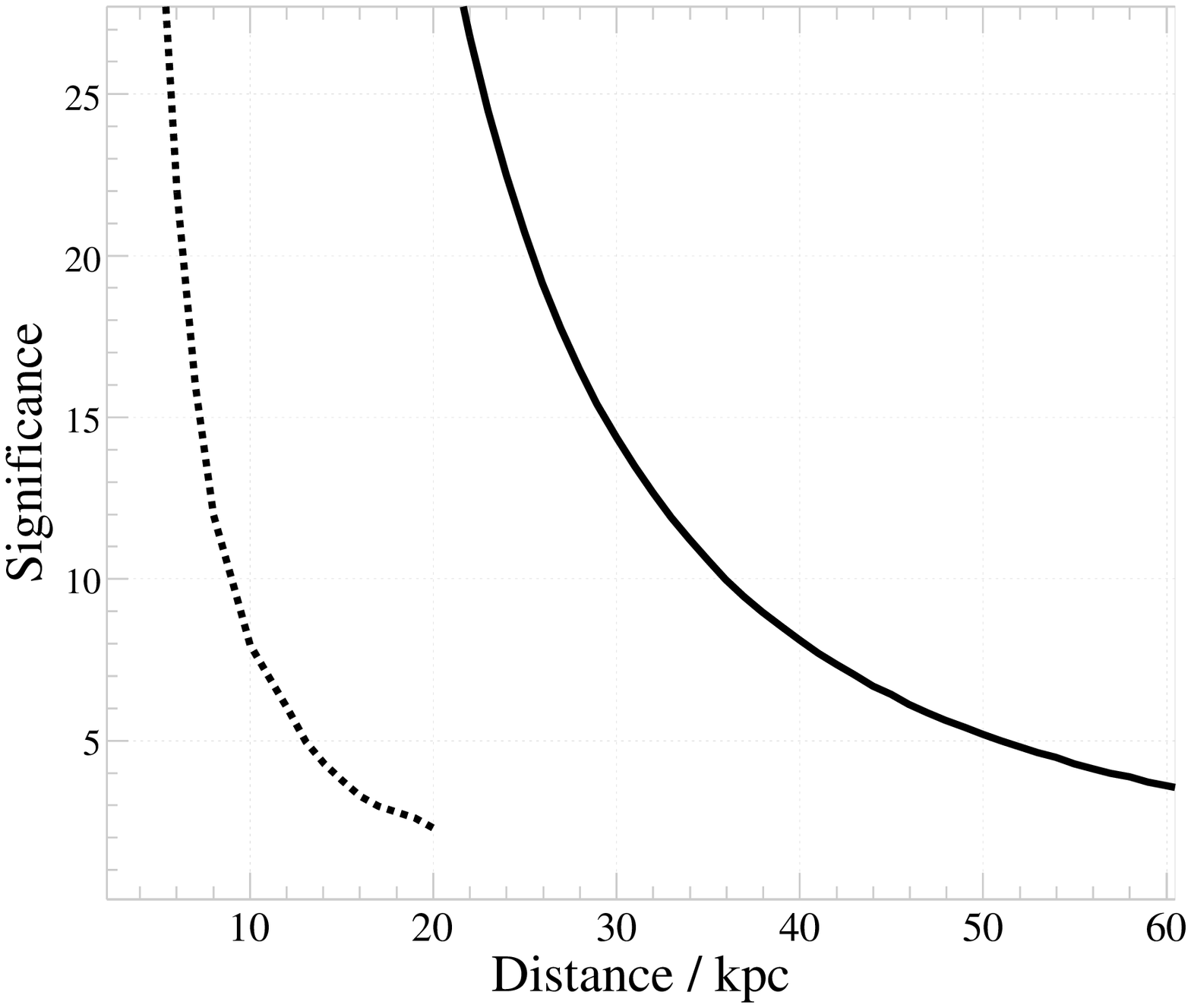}
	\end{center}
	\vspace{-0.4cm}
	\caption{Supernova detection ranges using $0.5\un{s}$ time frames}
	\footnotesize{Significances which supernovae conforming to the \emph{Lawrence-Livermore} model would
	cause as function of their distance. IceCube performance is described by the solid line, AMANDA 
	by the dotted line. Both are given for the $0.5\un{s}$ binning.}
	\label{fig::icecube_range}
\end{figure}

Figure~\ref{fig::ll_sn_7.5kpc} shows the expected signal in IceCube for a supernova at $7.5\un{kPc}$ conforming
to the \emph{Lawrence-Livermore} model with $\approx 10^6$ registered neutrinos in $15\un{s}$ and a statistical
accuracy of $0.1\%$ in the first $2 \un{s}$. Assuming $2 \cdot 10^4$ events in Super-Kamiokande (scaled 
from~\cite{Ikeda2007}), one arrives at an accuracy of $1\%$ in the same time frame. While IceCube can neither
determine the directions nor the energies of the neutrinos, it will provide worlds best statistical accuracy to 
follow details of the neutrino light curve. Its performance in this respect will be in the same order as proposed
megaton proton decay and supernova search experiments. As the signal is seen on top of background noise, 
the measurement accuracy drops rapidly with distance. Figure \ref{fig::icecube_range} shows the
significance at which IceCube and AMANDA would be able to detect supernovae.

\section{Deleptonization Peak and Neutrino Oscillations}

As mentioned before, the signal seen by AMANDA and IceCube is mostly dominated by electron anti-neutrinos
with a small contribution by electron neutrino scattering, thereby leading to a sensitivity which is
strongly dependent on the neutrino flavor and is thus sensitive to neutrino oscillations.

As the neutrinos pass varying levels of density within the supernova, the flux of electron and electron
anti-neutrinos $\Phi$ released is different from the initial flux $\Phi^0$ produced during the collapse:

\begin{eqnarray*}
\Phi_{\nu_e}		& = &	p\, \Phi^0_{\nu_e} + (1-p) \, \Phi^0_{\nu_x}
\quad , \\
\Phi_{\bar{\nu}_e} 	& = &	\bar{p} \, \Phi^0_{\bar\nu_e} + (1-\bar{p}) \, \Phi^0_{\bar\nu_x}
\quad .
\end{eqnarray*}

\noindent The survival probability $p / \bar p$ for the $\nu_e / \bar\nu_e$'s varies with the mass
hierarchy and the mixing angle $\theta_{13}$~\cite{Dighe2005}:

\begin{center}
	\begin{tabular}{lccc}
		neutrino oscillation parameters					& \vline & $p$		& $\bar p$ 	\\
		\hline
		$m_2^2 < m_3^2 \;,\; \sin^2\theta_{13}>10^{-3}		$	& \vline & $\approx0\%$	& $69\%$	\\
		$m_1^2 > m_3^2 < 0 \;,\; \sin^2\theta_{13}>10^{-3}	$	& \vline & $31\%$	& $\approx0\%$	\\
		any hierarchy, $\sin^2\theta_{13}<10^{-5}		$	& \vline & $31\%$	& $69\%$ 	\\
	\end{tabular}
	\label{tab::osci}
\end{center}

\noindent At the onset of the supernova neutrino burst during the prompt shock, a $\sim 10\un{ms}$ long
burst of electron neutrinos gets emitted when the neutron star forms. As the shape and rate of this burst
is roughly independent of the properties of the progenitor stars, it is considered as a standard candle, 
allowing one to determine neutrino properties without knowing details of the core collapse.  

\begin{figure}
	\begin{center}
		\includegraphics [width=0.47\textwidth]{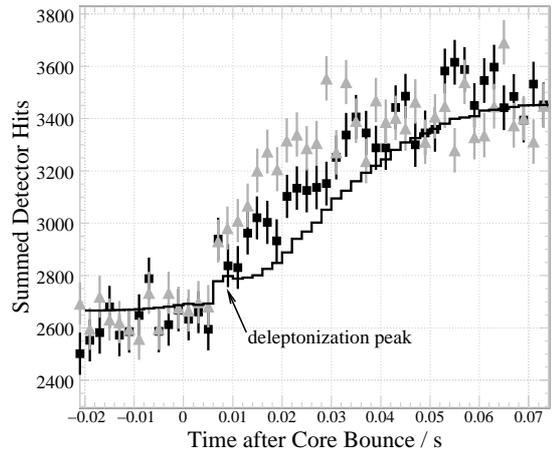}
	\end{center}
	\vspace{-0.4cm}
	\caption{Neutrino signals of a supernova at $7.5 \un{kpc}$ distance modified by oscillations as
	seen in IceCube.}
	\vspace{0.1cm}
	\footnotesize{The line shows the expectation without neutrino oscillations, the squares show
	the simulated signal for the normal mass hierarchy with $\sin^2\theta_{13}>10^{-3}$ 
	($\sin^2\theta_{13}<10^{-5}$ lies nearly on the same line) and the triangles show the signal for
	inverted mass hierarchy at $\sin^2\theta_{13}>10^{-3}$.} 
	\label{fig::sn_oscillation_scenarios_7.5kpc}
\end{figure}

The first $0.7\un{s}$ of a supernova signal were modeled after \cite{Kitaura2006}. 
Figure~\ref{fig::sn_oscillation_scenarios_7.5kpc} shows the expectation for a supernova at a distance of
$7.5\un{kpc}$ in the $2\un{ms}$ binning of IceCube. Due to statistics and the rising background from the
starting electron anti-neutrino signal, the identification of the neutronization burst is unlikely at this
distance. However, with a supernova at $7.5\un{kpc}$ one might be able to draw conclusions for the mass 
hierarchy, depending on the reliability of the models.

\section{Conclusions and Outlook}

With 59 strings and 3540 OMs installed, IceCube has reached $86\%$ of its final sensitivity for supernova 
detection. It now supersedes AMANDA in the SNEWS network. With the low energy extension DeepCore, IceCube 
gains 360 additional OMs with a $\sim 30\%$ higher quantum efficiency (rates $\sim 380\un{Hz}$). These
modules have not been considered in this paper.

\end{document}